\documentclass[copyright,creativecommons,noderivs,noncommercial]{eptcs}

\providecommand{\event}{B. Klin, P. Soboci\'nski (Eds.): \\
6th Workshop on Structural Operational Semantics (SOS'09)}
\providecommand{\volume}{19}
\providecommand{\anno}{2010}
\providecommand{\firstpage}{32}
\providecommand{\eid}{3}

\usepackage{color}
\usepackage{amsmath}
\usepackage{amssymb}
\usepackage{graphicx}

\newtheorem{seman}{Semantics}
\newtheorem{defn}{Definition}
\newtheorem{exmp}[defn]{Example}

\newtheorem{thm}[defn]{Theorem}

\newcommand{\sosrule}[2]{\frac{\raisebox{.7ex}{\normalsize{$#1$}}}
                        {\raisebox{-1.0ex}{\normalsize{$#2$}}}}
\newcommand{\trans}[1]{\,{\stackrel{{#1}}{\longrightarrow}}\,}
\newcommand{\emits}[1]{\uparrow^{#1}}
\newcommand{\ntrans}[1]{\,{\stackrel{{#1}}{\nrightarrow}}\,}

\newcommand{\tss}{\mathcal{T}}

\newcommand{\term}{\checkmark}

\newcommand{\DR}[1]{\ensuremath \mathbf{(#1)}}

\newcommand{\Terms}[1]{\mathbb{T}(#1)}
\newcommand{\CTerms}[1]{\mathbb{C}(#1)}

\newcommand{\vars}[1]{\mathit{vars}(#1)}
\newcommand{\conc}[1]{\mathit{conc}(#1)}
\newcommand{\prem}[1]{\mathit{prem}(#1)}

\newcommand{\subterms}[1]{\mathit{subterms}(#1)}

\newenvironment{bullets}{\begin{list}{$\bullet$}{\itemsep 0em\topsep 0em}\labelwidth 3mm}{\end{list}}

\definecolor{darkgreen}{rgb}{0,.6,0}

\newcommand{\ecomm}[1]{\texttt{#1}}

\newcommand{\einput}[1]{\ecomm{input}~#1;}
\newcommand{\eoutput}[1]{\ecomm{output}~#1;}
\newcommand{\enothing}{\ecomm{0}}
\newcommand{\eemit}[1]{\ecomm{emit}~#1}
\newcommand{\epause}{\ecomm{1}}
\newcommand{\epresent}[3]{\ecomm{pres}\ #1\ \ecomm{?}\ #2\ \ecomm{$\diamond$}\ #3\ \ecomm{end}}
\newcommand{\esuspend}[2]{\ecomm{susp}\ #1\ \ecomm{when}\ #2}
\newcommand{\eseq}[2]{#1\ \ecomm{;}\ #2}
\newcommand{\eloop}[1]{\ecomm{loop}\ #1\ \ecomm{end}}
\newcommand{\epar}[2]{#1\ \ecomm{||}\ #2}
\newcommand{\etrap}[2]{\ecomm{trap}\ #1\ \ecomm{in}\ #2\ \ecomm{end}}
\newcommand{\eexit}[1]{\ecomm{exit}\ #1}
\newcommand{\eencap}[2]{\ecomm{sign}\ #1\ \ecomm{in}\ #2\ \ecomm{end}}

\title{Causality in the Semantics of Esterel: Revisited}

\author{MohammadReza Mousavi
\institute{Department of Computer Science, Eindhoven University of Technology,
\\ {P.O.~Box~513}, NL-5600~MB~~Eindhoven, The Netherlands
}
}

\def\authorrunning{M.R.~Mousavi}
\def\titlerunning{Causality in the Semantics of Esterel}

\begin{document}
\maketitle

\bibliographystyle{alpha}

\begin{abstract}
We re-examine the challenges concerning causality in the semantics of Esterel
and show that they pertain to the known issues in the semantics of Structured Operational Semantics with negative premises.
We show that the solutions offered for the semantics of SOS also provide answers to the semantic challenges of Esterel and that
they satisfy the intuitive requirements set by the language designers.
\end{abstract}

\section{Introduction}

Esterel \cite{Berry99,Berry07} is an imperative synchronous language used for the specification
and programming of embedded systems. Esterel is based on the synchronous hypothesis, i.e.,
instantaneous reaction to signals and immediate propagation of signals in each time-instant.
The combination of the imperative programming style and the synchronous hypothesis in Esterel has led to semantic challenges addressed in the literature \cite{Berry92b,Berry99,Tini00,Tini01,Tardieu05,Potop02}.
In this paper, we present the main semantic challenge posed by Esterel, namely, the issue of causality. We show that it is reminiscent of
the semantic challenges \cite{Groote93,Bol96,vanGlabbeek04} in Structured Operational Semantics \cite{Aceto01} (esp. in the setting with negative premises; the same challenges were encountered before in logic programming \cite{Apt94}).
We then show that using the known solutions for the latter simplifies the presentation of the semantics of the former substantially and leads to the desired intuitive properties set forth by the language designers.

%\paragraph{Related work.}

The rest of this paper is organized as follows. In Section \ref{sec::cook}, we present a brief overview of the Esterel language and its intuitive semantics. Section \ref{sec::sos} introduces Structured Operational Semantics and notions of semantics and well-definedness associated with SOS specifications.
Section \ref{sec::bridge} connects these two worlds by first presenting an SOS specification for Esterel and
then studying the notions of semantics and well-definedness for the given specification. There,
we show that certain notions of semantics for SOS formalize the intuitive criteria given by the language designers.
Section \ref{sec::conc} concludes the paper and presents directions for future research.

\begin{figure}
\[
\begin{array}{lcl}
 p,q &::=& \enothing ~|~ \eemit{s} ~|~ \epresent{s}{p}{q} ~|~ \\
     &   & \eseq{p}{q} ~|~ \epar{p}{q} ~|~  \eencap{s}{p} ~|~ \\
     &   & \epause ~|~ \esuspend{p}{s} ~|~ \etrap{t}{p} ~|~ \eexit{t} ~|~ \eloop{p}
\end{array}
\]
\caption{The Abstract Syntax of Esterel \label{fig::grammarEsterel}}
\end{figure}

\section{Esterel and Its Semantics: A Cook's Tour \label{sec::cook}}
The abstract syntax of Esterel is given by the grammar in Figure \ref{fig::grammarEsterel}.

A short introduction to the intuitive semantics of each of these constructs follows.
In this grammar, $\enothing$ stands for the terminated process.
Emitting signal $s$ is denoted by $\eemit{s}$,
which is instantaneously visible to all parts of the system (and may in turn cause more signals to be emitted).
Reacting to present and absent signals is done via the if-then-else construct $\epresent{s}{p}{q}$,
where if $s$ is currently present (emitted by some other part of the system),
$p$ is executed, otherwise if $s$ is absent  $q$ is executed.
The combination of synchronous assumption, i.e., instantaneous propagation of signals,
and checking for absence/presence of signals leads to semantic complications, presented shortly.
Parallel composition of $p$ and $q$ is denoted by $\epar{p}{q}$.
Process $\eencap{s}{p}$ encapsulates $s$ in $p$, i.e., declares $s$ local to $p$.
Another way of reacting to signals is by using the suspend construct  $\esuspend{p}{s}$, which initially acts as $p$,
but after one synchronous round will stop $p$ as soon as signal $s$ is emitted (suspension may happen after a number of rounds).
Process $\epause$ stands for a process that passes one unit of time and then terminates.
One can define traps (exit points, exception handlers) to which a program can jump to by $\etrap{t}{p}$. The actual jump (raising the exception) is performed by executing $\eexit{t}$.
A program can engage in a loop by means of $\eloop{p}$ (it can either keep on executing in the loop or exit the loop using $\eexit{t}$).

%As a syntactic sugar, we may write $\eemit{\overrightarrow{s}}$ (similarly, $\eencap{\overrightarrow{s}}{p}$) for a sequence of signals, which stands for $\enothing$ if $\overrightarrow{s}$ is the empty sequence and $\eemit{s_0} ; \ldots ; \eemit{s_n} $ if $\overrightarrow{s} = s_0, \ldots, s_n$.

An Esterel program is usually suffixed by a header declaring input and output signals.
The syntax of this header is of the form $\einput{i}  \eoutput {o}  $
and we assume that the set of input and output variables in a program is disjoint from the set of its local signals.
Moreover, to unclutter the syntax, we assume fixed sets $\iota$, $\omega$ and $\lambda$, respectively, of input, output and local variables.
We pick typical members  $i, i', i_0, \ldots \in \iota$, $o, o', o_0, \ldots \in \omega$ and $s, s', s_0 \in \lambda$.
This way, one does not need to consider the input and out declaration anymore since input and output (and local) variables are recognized by their names.
In some cases output and local variables can be treated uniformly, in which case we denote them by $x, x', x_0 \in \omega \cup \lambda$.

To study the semantic challenge concerning causality it suffices for us to look at the first two rows of our grammar.
The other constructs, e.g., traps and time passing, are semantically interesting on their own
but are treated satisfactorily in the literature and are orthogonal to the causality problems addressed here.
Hence, in the remainder, we focus on the subset of Esterel given in the first two lines of our grammar
and only in passing mention how to include time and traps in our presented semantics.

A causality relation between events $s$ and $s'$ (signals in this case), means that the presence and absence of $s$
directly influences the presence or absence of $s'$. For example, consider the following Esterel program:

\begin{enumerate}
\item[P0] {\eseq{\epresent{i}{\eemit{$\mathrm{s}$}}{\enothing}}{\epresent{$\mathrm{s}$}{\enothing}{\eemit{$\mathrm{o}$}}}}
\end{enumerate}

\noindent
In the above program, there is a causality chain starting from the input variable $i$ to the local variable $s$ and from $s$ to the output variable $o$, namely the presence of $i$ determines the presence of $s$ and eventually leads to the absence of $o$, while the absence of $s$ (caused by the absence of $i$), determines the presence of $o$. Using the syntax of Esterel one can easily write programs with cyclic
dependencies  (e.g., $s$ is present if and only if $s$ is present) or even worse, cyclic dependencies of a paradoxical nature (e.g., $s$ is present if and only if $s$ is absent). To illustrate these issues in Esterel, consider the following simple programs, which are all due to \cite{Berry99}.
These programs are canonical examples of different issues concerning causality in Esterel programs.

\begin{enumerate}
\item[P1] %\eencap{s}
{\epresent{s}{\eemit{s}}{\enothing}}

Program P1 relies on the presence of $s$ in order to emit signal $s$.
The logical semantics of Esterel \emph{rejects} this program on the ground that it has two ``models''.
The first one is by assuming that $s$ is present, which leads to a justification of this assumption by emitting $s$.
The other one is by assuming that $s$ is absent, which is supported by that $\enothing$  does not emit (denies emitting) signal $s$.

In each synchronous round, the ``model'' of an Esterel program is defined by a
\emph{global status}, which defines the status (presence/absence) of signals in this round.
A global status of a program is called \emph{coherent}
when the presence/absence of signals are determined consistently by the emit statements in the program \cite{Berry99}:

\begin{quote}
The global status of a program is \emph{logically coherent} iff at least one \eemit{} statement is executed for each signal assumed present and no \eemit{} statement is executed for each signal  assumed absent.
\end{quote}

For example, program P1 has two logically coherent global statuses, namely presence of $s$ and absence of $s$, as motivated above.
The basis for rejecting program P1 is called ``logical determinism'' and is defined as follows \cite{Berry99}.

\begin{quote}
A program is logically deterministic if it has at most one logically coherent global status.
\end{quote}

\item[P2] %\eencap{s}
{\epresent{s}{\enothing}{\eemit{s}}}

Program P2 relies on the absence of $s$ in order to emit signal $s$.
According to the logical semantics of Esterel, the above-given program has no logically coherent global status.
Assuming that $s$ is absent leads to $s$ being emitted and hence, incoherency.
Likewise, assuming that $s$ is present requires emission of $s$, which is only justified when $s$ is absent.

The basis for rejecting program P2 is called ``logical reactivity'' and is defined as follows \cite{Berry99}.

\begin{quote}
A program is logically reactive if it has at least one logically coherent global status.
\end{quote}

The conjunction of logical determinism and logical reactivity is called logical coherency and is the main well-definedness criterion for the
\emph{logical semantics} of Esterel.

\item[P3] %\eencap{s}
{\epresent{s}{\eemit{s}}{\eemit{s}}}

The program above has only one logically coherent global status, namely that $s$ is present.
This global status is also coherent since assuming the presence of $s$ leads to emitting it and moreover,
it is not logically coherent to assume the absence of $s$, because it leads to its emission.
Hence, as far as logical coherency is concerned this program is accepted and the logical semantics
defines the semantics sketched above for this program.

However, the semantics of Esterel used for its compiler, called the \emph{constructive semantics} \cite{Berry99,Berry07},
has further constraints which lead to the rejection of the above program.
In this paper, we consider the issue of causality in both variants of the semantics
and hence, also study the issue of constructiveness defined below.

\begin{quote}
A program is \emph{constructive}, if for each signal, it either proves its presence (must emit the signal) or proves its absence (cannot emit it).
\end{quote}

Program P3 is rejected by the above criterion since it can neither prove the emission of $s$ (its only possible proof is cyclic since relies on the assumption that $s$ is emitted), nor can it coherently prove its absence, since to prove the absence of $s$ it should prove that neither of the two
emit statements can be executed, thus it should prove that $s$ can neither be present nor absent.

\item[P4] %\eencap{$\mathrm{s_0,s_1}$}
{\epar{\hspace{0.2cm} \epresent{$\mathrm{s_0}$}{\eemit{$\mathrm{s_0}$}}{\enothing}}
{\\ \hspace{0.2cm} \epresent{$\mathrm{s_0}$}{\epresent{$\mathrm{s_1}$}{\enothing}{\eemit{$\mathrm{s_1}$}}}{\enothing}} \\}

Note that P4 is logically coherent, since its only logically coherent global status is
that both $s_0$ and $s_1$ are absent.
To check its constructiveness, let us focus on the emission of $s_0$. It definitely does not have to emit $s_0$, since the only
reason for emitting $s_0$ is the \eemit{$\mathrm{s_0}$} statement in the left-hand side of the parallel composition, which is guarded by the check
on the presence of $s_0$. Hence, the only proof for emitting $s_0$ is cyclic.
But it can potentially emit $s_0$ (because it contains an  \eemit{$\mathrm{s_0}$}  statement) and the only way to make sure that $s_0$ cannot be emitted is to prove that the guard for \eemit{$\mathrm{s_0}$}  never becomes true, i.e., we need again to show that $s_0$ cannot be emitted, which is also a cyclic reasoning.
Hence, we conclude that P4 is not constructive because it neither must emit $s_0$, nor it can deny its emission.

\item[P5] %\eencap{$\mathrm{s_0,s_1}$}
{\eseq{\epresent{$\mathrm{s_0}$}{\eemit{$\mathrm{s_1}$}}{\enothing}}{\eemit{$\mathrm{s_0}$}}}

The above program is logically coherent and its unique logically coherent global status is that both $s_0$ and $s_1$ are present.
However, it is again rejected by the constructive semantics of Esterel.
The reason is that in order to reach the emit statement for $s_0$, we should first make sure that the first statement has a well-defined semantics in this context,
i.e., it either takes the if branch or the else branch and then terminates. However, giving a constructive proof for the transition of the conditional requires a constructive proof for the emission of $s_0$. This is another instance of the cyclic proof phenomenon rejected by the constructive semantics.

\end{enumerate}

\section{Structured Operational Semantics\label{sec::sos}}
Structural Operational Semantics (SOS) was originally proposed by Plotkin \cite{Plotkin04b} as a syntax-directed and compositional way of defining semantics.
Gradually, SOS has gained popularity and by now has become a de facto standard in defining operational semantics.
This popularity has called for a richer syntax for SOS deduction rules and  thus, in some applications, SOS deduction rules lost their structural, i.e., inductive, nature.
Some authors then decided to use the same acronym for Structured Operational Semantics \cite{Aceto01,Groote92}.
With the richer syntax of SOS rules, one can write deduction rules whose meaning is not clear any more.

\begin{exmp}\label{ex::cyclicRules}
Examples of cyclic rules are the deduction rules $\DR{r1}$ and $\DR{r2}$ given below.

\[
\DR{r1}\sosrule{p \trans{s} p}{p \trans{s} p} \quad \quad \DR{r2}\sosrule{p \ntrans{s} }{p \trans{s} p}
\]

The reader may already note the curious similarity between program P1 and deduction rule $\DR{r1}$ on one hand and
program P2 and deduction rule $\DR{r2}$ on the other hand.
Moreover, program P3 resembles the combination of $\DR{r1}$ and $\DR{r2}$.
These similarities materialize as formal definitions in the remainder of this paper.
\end{exmp}

To formalize the syntax and semantics of SOS, we first formalize the concepts of formulae and (transition) formulae.

\begin{defn}[Signature and (sub)terms]
    We let $V$ represent an infinite set of variables.
    A \emph{signature} $\Sigma$ is a set of function symbols (operators), each with a fixed arity. An operator with arity zero is called a
    \emph{constant}. We define the set $\Terms\Sigma$ of \emph{terms} over $\Sigma$ as the smallest set satisfying the following constraints.
    \begin{bullets}
        \item A variable $x\in V$ is a term.
        \item If $f\in \Sigma$ has arity $n$ and $t_1,\dots,t_n$ are terms, then $f(t_1,\dots,t_n)$ is a term.
    \end{bullets}
    We write $t_1 \equiv t_2$ if $t_1$ and $t_2$ are syntactically equal.
    The function $vars : \Terms\Sigma \rightarrow 2^V$ gives the set of variables appearing in a term.
    The set $\CTerms\Sigma \subseteq \Terms\Sigma$ is the set of \emph{closed terms}, i.e., terms that contain no variables.
    A \emph{substitution} $\sigma$ is a function of type $V \rightarrow \Terms\Sigma$. We extend the domain of substitutions to terms
    homomorphically. If the range of a substitution lies in $\CTerms\Sigma$, we say that it is a \emph{closing substitution}.

    A term $s$ is considered a \emph{subterm} of itself;
    if $s$ is a subterm of $t_i$, then $s$ is also a subterm of $f(t_0, \ldots, t_i, \ldots, t_{n-1})$, for each $s, t_i \in \Terms{\Sigma}$, $0 \leq i < n$, and n-ary $f \in \Sigma$. The set of subterms of a term $t$ are denoted by $\subterms{t}$.
\end{defn}

Next, we formalize the syntax of SOS in terms of Transition System Specifications.

\begin{defn}[Transition System Specifications (TSS)]
    A \emph{transition system specification} is a triplet\\ $(\Sigma, L, D)$ where
    \begin{bullets}
        \item $\Sigma$ is a signature.
        \item $L$ is a set of labels. If $l \in L$, and $t,t'\in \Terms\Sigma$
              we say that $t \trans{l} t'$ is a \emph{positive formula} and
              $t \ntrans{l}$ (also denoted by $\neg t \trans{l}$) and $t \ntrans{l} t'$ are \emph{negative formulae}. A formula, typically denoted by $\phi$, $\psi$, $\phi'$, $\phi_i$, $\ldots$
              is either a negative formula or a positive one.
        \item $D$ is a set of \emph{deduction rules}, i.e., tuples of the form $(\Phi,\phi)$ where $\Phi$ is a set of
              formulae and $\phi$ is a positive formula. We call the formulae contained in $\Phi$ the \emph{premises} of the rule and $\phi$ the
              \emph{conclusion}.
    \end{bullets}
    We write $\vars{r}$ to denote the set of variables appearing in a deduction rule $\DR{r}$.
    We say a formula is \emph{closed} if all of its terms are closed. Substitutions are also extended to formulae
    and sets of formulae in the natural way.
\end{defn}

A deduction rule $(\Phi,\phi)$ is typically written as $\frac{\Phi}{\phi}$. For a deduction rule $r$, we write $\conc{r}$ to denote its conclusion and $\prem{r}$ to denote its premises. A set of positive closed formulae is called a \emph{transition relation}.  Given a transition relation $T$, $L'$-labeled transitions of closed term $p$, denoted by $T \downarrow (p,L')$ is the subset of $T$ containing all formulae in $T$ that have $p$ as their source and some $l \in L'$ as their label.
A TSS is supposed to define a transition relation but for the TSSs such as those given by deduction rules $\DR{r0}$ and $\DR{r1}$, it is not clear what the associated transition relation is. Several proposals are given in the literature, of which \cite{vanGlabbeek04} gives a comprehensive overview and comparison.
In this paper, we shall use some of these proposals to define the semantics of Esterel.
In order to facilitate the presentation of these proposals, we need two auxiliary definitions, namely contradiction and contingency, which are given below.

%\begin{defn}[Transition Relations and Reachability]
%Given a transition relation $T$, reachability relation in $T$ between terms is the smallest relation satisfying the following constraints, for each
%$q, q'\in \CTerms{\Sigma}$,
%\begin{enumerate}
%\item $p$ is reachable from $p$ in $T$,  and
%\item if $q$ is areachable from $p$ and $q \trans{l} q' \in T$, then $q'$ is reachable from $p$.
%\end{enumerate}
%\end{defn}

\begin{defn}[Contradiction and Consistency]
\hspace*{-2pt}Formula $t \trans{l} t'$ is said to \emph{contradict} both $t \ntrans{l}$ and $t \ntrans{l} t'$, and vice versa.
%For two sets $\Phi$ and $\Psi$ of formulae, $\Phi$ \emph{contradicts} $\Psi$, denoted by $\Phi \nvDash \Psi$, when there is a $\phi \in \Phi$ that contradicts a $\psi \in \Psi$.
$\Phi$ is \emph{consistent} w.r.t.\ $\Psi$, denoted by $\Phi \vDash \Psi$, when for each positive formula $\psi \in \Psi$, it holds that $\psi \in \Phi$ and
for each negative formula $\psi \in \Psi$, there is no $\phi \in \Phi$ such that $\phi$ contradicts $\psi$.
\end{defn}

In the remainder, we only use negative formulae of the form $t \ntrans{l}$ in our specifications.
%It immediately follows from the above definition that contradiction and contingency are symmetric relations on (sets of) formulae.
We now have all the necessary ingredients to present different proposals for the semantics of TSSs.
The first proposal is the following notion of supported model, which is a slight modification of the definition in \cite{vanGlabbeek04}
(restricting it to particular sets of terms and labels).

\begin{defn}[Supported Model\label{def::supported}]
Given A TSS, a transition relation $T$ is a supported model for a set $P \subseteq \CTerms{\Sigma}$ of closed terms and a set $L' \subseteq L$ of labels, when
\begin{enumerate}
  \item for each $q, q' \in \Terms{\Sigma}$ and $l \in L$ if $q \trans{l} q' \in T$, then there exists a deduction rule $\sosrule{\Phi}{\phi}$ and a substitution $\sigma$ such that $\sigma(\phi) = q \trans{l} q'$ and $T \vDash \Phi$, and
\item for each $p \in P$ and $l \in L'$, $p' \in \Terms{\Sigma}$, if  there exists a deduction rule $\sosrule{\Phi}{\phi}$ and a substitution $\sigma$ such that $\sigma(\phi) = p \trans{l} p'$ and $T \vDash \Phi$, then $p \trans{l} p' \in T$.
\end{enumerate}

A transition relation $T$ is a supported model for a TSS when it is a supported model for $\CTerms{\Sigma}$ and $L$.
\end{defn}

Note that in the above definition and throughout the rest of the paper, we only consider the ``immediate transitions'' of $p$ as its semantics.
One can adapt the above definitions (and the subsequent ones) to consider the ``transition system'' associated with $p$ as its semantics.
For the subset of Esterel considered in this paper, these two notions lead to the same conclusion concerning the well-definedness and the semantics of a program.

\begin{seman}[Unique Supported Model Semantics\label{sem::uniqueSupported}]
Given a set $P \in \CTerms{\Sigma}$  of closed terms and a set $L' \subseteq L$ of labels a TSS is \emph{meaningful} w.r.t.\ $P$ and $L'$ when it has a unique supported model for $P$ and $L'$; the \emph{transition system} associated with $P$ and $L'$ is the unique supported model for $P$ and $L'$.
A TSS is meaningful when it has a unique supported model; the transition relation associated with a TSS is its unique supported model.
\end{seman}

To illustrate these concepts, we give a few simple TSSs and study their supported models.

\begin{exmp}\label{ex::supported}
Consider the deduction rules given in Example \ref{ex::cyclicRules}.

Consider the TSS comprising only deduction rule $\DR{r1}$.
This TSS is not meaningful (w.r.t.\ $\{p\}$ and $\{s\}$) according to Semantics \ref{sem::uniqueSupported} because it has two supported models, namely $\emptyset$ and $\{p \trans{s} p\}$.

Also according to Semantics \ref{sem::uniqueSupported},  the TSS comprising only deduction rule $\DR{r2}$ is not meaningful  (w.r.t.\ $\{p\}$ and $\{s\}$)  either, because it has no supported model. Particularly, $T = \emptyset$ is not a supported model because it follows from the right-to-left implication of Definition \ref{def::supported} that $p \trans{s} p \in T$.
$T = \{p \trans{s} p \}$ is not a supported model either since the only deduction rule providing a reason for $p \trans{s} p \in T$ is $\DR{r2}$ but
it does not hold that $T \vDash p0 \ntrans{s}$.

The TSS comprising both $\DR{r1}$ and $\DR{r2}$ is indeed meaningful and its associated transition relation is $T = \{ p \trans{s} p\}$.
Transition relation $T$ is indeed  a supported model since $\DR{r1}$ now provides a reason for $p \trans{s} p \in T$.
Moreover $T' = \emptyset$ is not a supported model for this TSS because it then follows from $\DR{r2}$ and
the right-to-left implication of Semantics \ref{sem::uniqueSupported} that $p \trans{s} p \in T'$.

If one takes the transition system of a program as a formalization of its global state,
then the TSS comprising of deduction rule $\DR{r1}$ is rejected because it has no coherent global state and
the TSS with only $\DR{r2}$ is rejected because it does equivocally define a coherent global state.
\end{exmp}

This suggests that Semantics \ref{sem::uniqueSupported} provides a suitable formalization for logical coherency.
Next, we give a formalization of constructiveness in terms of supported proofs and denials.

\begin{defn}[Supported Proofs\label{def::supportedProof}]
A TSS $\tss$ provides a supported proof for a formula  $\phi$, denoted by $\tss \vdash_s \phi$, when there is a well-founded upwardly branching tree with formulae as nodes and of which
\begin{itemize}
\item the root is labelled by $\phi$;
\item if a node is labelled by a positive formula $\psi$ and the nodes above it form the set $K$ then $\frac{K}{\psi}$ is an instance of a deduction rule in $\tss$.
\item if a node is labelled by a negative formula $\psi$, and the nodes above it form the set $K$,  then for each instance of a deduction rule $\frac{K_i}{\psi_i}$ in $\tss$ such that $\psi_i$ contradicts $\psi$, there exists a formula $\psi'_i \in K$ contradicting a formula in $K_i$.
\end{itemize}
\end{defn}

\begin{seman}[S-Complete Semantics\label{sem::sComplete}]
A TSS is $s$-complete for a set of closed terms $P$ when for each formula $\phi$
with a $p \in P$ as its source, either $\phi$ or a formula contradicting it has a supported proof.
A TSS is $s$-complete when it is $s$-complete for the set $\CTerms{\Sigma}$ of all closed terms and
its transition relation is the set of positive formulae, for which it provides supported proofs.
\end{seman}

The following theorem is taken from \cite{vanGlabbeek04}, which shows that constructiveness is indeed stronger than logical coherency.

\begin{thm}\label{th::completeImpliesUnique}
A program (TSS) is $s$-complete only if it is meaningful according to Semantics \ref{sem::uniqueSupported} (has a unique supported model) and its associated transition system (unique supported model) coincides with the set of all positive formulae with a supported proof.
\end{thm}

Another useful property of supported proofs is their consistency \cite{vanGlabbeek04}, stated below.

\begin{thm}\label{th::supportedConsistency}
The notion of supported proof is consistent, i.e., for each formula $\phi$ with a supported proof, its negation does not have a supported proof.
\end{thm}

Next, we re-examine the TSS of Example \ref{ex::supported} using our new notion of semantics.

\begin{exmp}
The two TSSs comprising only $\DR{r1}$ and only $\DR{r2}$ are both rejected by Semantics \ref{sem::sComplete}, as well,
since neither $p \trans{s} p$, nor $p \ntrans{s}$ can be proven from either of them.
(This is also an immediate consequence of Theorem \ref{th::completeImpliesUnique}.)

In the case of the TSS comprising only $\DR{r1}$, any attempt to build a supported proof for $p \trans{s} p$ has the same formula as its premise.
Moreover, $p \ntrans{s}$ cannot be proven because its proof tree should prove a negation of a premise of $\DR{r1}$, i.e., again $p \ntrans{s}$.
In other words, both $p \trans{s} p$ and $p \ntrans{s}$ only have cyclic, and thus unsupported, proofs.

Similarly, in the case of the TSS comprising only $\DR{r2}$, neither $p \trans{s} p$, nor $p \ntrans{s}$ have a supported proof.

Consider the TSS comprising both $\DR{r1}$ and $\DR{r2}$; it does have a unique supported model $T = \{p \trans{s} p\}$ but
it is \emph{not} $s$-complete and is thus rejected by Semantics \ref{sem::sComplete}.
Any proof for $p \trans{s} p$ or its negation leads to a cycle, i.e., repeating the node below in the node above, and are thus not supported.

Again drawing an analogy with Esterel programs, Semantics \ref{sem::sComplete} requires the existence of a ``constructive'' (supported) proof
for presence/absence of signals and thus rejects a program which uses both \emph{possibilities} for a signal in order to establish its own presence.
\end{exmp}

\section{Structured Operational Semantics for Esterel\label{sec::bridge}}
Our semantic specification of Esterel is presented in Figures \ref{fig::sosEmission} and \ref{fig::sosEsterel}.
The state of the SOS comprises the syntax of the program currently being executed (defined by the grammar in Figure \ref{fig::grammarEsterel}).
The semantics is supposed to define two predicate, $p \term_{I, c}$, $p \emits{I, c, s}$, respectively, where the former  means that $p$ terminates
with input evaluation $I$ and under context $c$ (if $p$ is part of program $c$), and  the latter means that $p$ emits signal $s$ (in the present time-instant) under the same assumptions.
(A predicate formula can be formally interpreted as a transition formula with a dummy right-hand-side; in our case one can take $\enothing$ to be the dummy target of all predicate formulae, i.e., read $p \emits{I, c, s}$ and $p \term_{I,c}$ as $p \emits{I, c, s} \enothing$ and $p \term_{I, c} \enothing$, respectively.)
In addition to the two predicates, the semantics is supposed to define a transition relation of the form $p \trans{I, c, s} p'$, which denotes that program $p$ emits signal $s$ under input evaluation $I$ and context $c$.
Next, we briefly describe the deduction rules in Figures \ref{fig::sosEmission} and \ref{fig::sosEsterel}
and then show how they formalize the intuitive properties of Esterel programs discussed before.
In all labels (of predicates and transitions) of Figures \ref{fig::sosEmission} and \ref{fig::sosEsterel},
$I \subseteq \{i^+, i^- \mid i \in \iota\}$ such that for each $i \in \iota$, either $i^- \in I$ or $i^+ \in \iota$ (but not both),
$c \in \CTerms{\Sigma}$, $i \in \iota$, $x \in \omega \cup \lambda$ and $s, s', s'' \in \lambda$.

\begin{figure}
\[
\begin{array}{c}
\DR{e0} \sosrule{}{\eemit{x} \emits{I, c, x}} \quad\quad
\DR{s0} \sosrule{p \emits{I, c, x}}
{\eseq{p}{q} \emits{I, c, x}} \quad \quad
\DR{s1} \sosrule{p \term_{I, c} \quad q \emits{I, c,x}}{\eseq{p}{q} \emits{I, c, x}} \quad \quad
\DR{s2} \sosrule{p \trans{I, c, x'} p' \quad p' \term_{I,c} \quad q \emits{I, c, x}}{\eseq{p}{q} \emits{I, c, x}}
\\ \\
\DR{p0} \sosrule{p \emits{I, c, x}}
       {\epar{p}{q} \emits{I, c, x}} \quad \hspace*{-8pt} \quad
\DR{p1} \sosrule{q \emits{I, c, x}}
       {\epar{p}{q} \emits{I, c, x}}
\quad \hspace*{-8pt} \quad
\DR{f0}
\sosrule{c \emits{I, c, s} \quad p \emits{I, c,x}}{\epresent{s}{p}{q} \emits{I, c,x}}
\quad \hspace*{-8pt} \quad
\DR{f1}
\sosrule{\neg c \emits{I, c, s} \quad q \emits{I, c, x}}{\epresent{s}{p}{q} \emits{I, c, x}}
\\ \\
\DR{f2}
\sosrule{i^+ \in I \quad p \emits{I, c,x}}{\epresent{i}{p}{q} \emits{I, c,x}}
\quad \quad
\DR{f3}
\sosrule{i^- \in I \quad q \emits{I, c, x}}{\epresent{i}{p}{q} \emits{I, c, x}}
\\ \\
\DR{en0} \sosrule{p[s''/s] \emits{I, c, s'}}{\eencap{s}{p} \emits{I, c,s'[s/s'']}} \quad s''\  \mbox{fresh in } p \mbox{ and } r \quad \quad
%\\ \\
%\DR{pr}\sosrule{p \emits{\eseq{\eemit{\overrightarrow{j}}}{p}, s}}{\einput{i}\eoutput{o} p \pemits{s}{\overrightarrow{j}}} \overrightarrow{j} \subseteq \overrightarrow{i}, s \notin \overrightarrow{i}
\end{array}
\]
\caption{Structured Operational Semantics for Esterel (Part I: Signal Emission)\label{fig::sosEmission}}
\end{figure}

In Figure \ref{fig::sosEmission}, $\DR{e0}$ states that $\eemit{s}$ can emit signal $s$ under any arbitrary input evaluation and context.
Deduction rules $\DR{s0}$, $\DR{s1}$ and $\DR{s2}$ describe when a sequential composition emits a signal, namely,
when either the first component of the composition emits it, or when the first component terminates (possibly after a transition)
and the second component emits the signal.

The notions of termination and transition are defined in Figure \ref{fig::sosEsterel}.
A parallel composition emits a signal if one of its components emits the signal, which is captured by deduction rules $\DR{p0}$ and $\DR{p1}$.
An if-then-else constructs emits a signal, if either, according to deduction rule $\DR{f0}$,
the local signal in its condition is emitted and the if-branch emits the signal or,
according to deduction rule $\DR{f1}$, the local signal in the condition cannot be emitted and the else-branch is taken.
Deduction rules $\DR{f2}$ and $\DR{f3}$ take care of the case where the condition is an input signal.
In such cases, the condition is checked against the given input evaluation.
A program $p$ with a local signal $s$ can emit a signal $s'$, if $p$ with a fresh signal $s''$ substituted for $s$ can emit
$s'$ (but if $s'$ is $s$, then $p$ should be able to emit $s''$).
%Finally, a program with a (possibly empty) list of inputs $\overrightarrow{i}$ can emit a signal $s$, if it can emit $s$ while some of its inputs are  ssumed to be present.

\begin{figure}
\[
\begin{array}{c}
\DR{nil} \sosrule{}{\enothing \term_{I,c}} \quad \quad  \DR{em} \sosrule{}{\eemit{x} \trans{I, c, x} \enothing}
\\ \\
%\sosrule{}{\epause \ttrans{} \term} \quad \quad
%\sosrule{p \trans{p'', s} p'}{\eseq{p}{q} \trans{p'', s} p'}\quad
%\sosrule{p \trans{c, s} \term_k}{\eseq{p}{q} \trans{c, s'} q'}\quad \quad
\DR{seq0} \sosrule{p \trans{I, c, x} p' \quad p' \term_{I,c} \quad q \trans{I, c, x'} q'}
{\eseq{p}{q} \trans{I, c, x} q'} \quad \quad
\DR{seq1} \sosrule{p \trans{I, c, x} p' \quad p' \term_{I, c} \quad q \trans{I, c, x'} q'}{\eseq{p}{q} \trans{I, c, x'} q'}\\\\
%\DR{seq2} \sosrule{p \trans{c, s} p' \quad \neg p' \term_0}{\eseq{p}{q} \trans{c, s} \eseq{p'}{q}}  \\ \\
\DR{seq2} \sosrule{p \term_{I, c} \quad q \trans{I, c, x} q'}{\eseq{p}{q} \trans{I, c, x} q'}  \quad
\DR{seq3} \sosrule{p \trans{I, c, x} p' \quad p' \term_{I, c} \quad q \term_{I, c}}{\eseq{p}{q} \trans{I, c, x} p'}  \quad
\DR{seq4} \sosrule{p \term_{I, c} \quad q \term_{I, c}}{\eseq{p}{q} \term_{I, c}}
\\ \\
\DR{par0} \sosrule{p \trans{I, c, x} p' \quad q \trans{I, c, x'} q'}
       {\epar{p}{q} \trans{I, c, x} \epar{p'}{q'}} \quad
\DR{par1} \sosrule{p \trans{I, c, x} p' \quad q \trans{I, c, x'} q'}
       {\epar{p}{q} \trans{I, c, x'} \epar{p'}{q'}} \\ \\
\DR{par2} \sosrule{p \term_{I, c} \quad q \trans{I, c, x} q'}
       {\epar{p}{q} \trans{I, c, x} q'} \quad
\DR{par3} \sosrule{p \trans{I, c, x} p' \quad q \term_{I, c} }
       {\epar{p}{q} \trans{I, c, x} p'} \quad
\DR{par4} \sosrule{p \term_{I, c} \quad q \term_{I, c}}
       {\epar{p}{q} \term_{I, c}}
       \\ \\
\DR{if0}
\sosrule{c \emits{I, c,s} \quad p \trans{I, c,x} p'}{\epresent{s}{p}{q} \trans{I, c, x} p'}
\quad \quad
\DR{if1}
\sosrule{\neg c \emits{I, c, s} \quad q \trans{I, c, x} q'}{\epresent{s}{p}{q} \trans{I, c, x} q'}
\\ \\
\DR{if2}
\sosrule{i^+ \in I \quad p \trans{I, c, x} p'}{\epresent{i}{p}{q} \trans{I, c, x} p'}
\quad \quad
\DR{if3}
\sosrule{i^- \in I \quad q \trans{I, c, x} q'}{\epresent{i}{p}{q} \trans{I, c, x} q'}
\\ \\
\DR{if4}
\sosrule{c \emits{I, c, s}  \quad p \term_{I,c}}{\epresent{s}{p}{q} \term_{I,c}}
\quad \quad
\DR{if5}
\sosrule{\neg c \emits{I, c, s} \quad q \term_{I,c}}{\epresent{s}{p}{q} \term_{I,c}}
\\ \\
\DR{if6}
\sosrule{i^- \in I \quad p \term_{I,c}}{\epresent{i}{p}{q} \term_{I,c}}
\quad \quad
\DR{if7}
\sosrule{i^+ \in I \quad q \term_{I,c}}{\epresent{i}{p}{q} \term_{I,c}}
\\ \\
\DR{enc0} \sosrule{p[s''/s] \trans{c, s'} p'}{\eencap{s}{p} \trans{c,s'[s/s'']} \eencap{s}{p'[s/s'']}} \quad
\DR{enc1} \sosrule{p[s''/s] \term_c}{\eencap{s}{p} \term_{I,c}}
\\\\
s''\  \mbox{fresh in } p \mbox{ and } r
%\\ \\
%\DR{prog}\sosrule{p \trans{\eseq{\eemit{\overrightarrow{j}}}{p}, s''} p'}{\einput{i}\eoutput{o} p \ttrans{s''}{\overrightarrow{j}} p'} \quad
%\DR{prog-term} \sosrule{p \term_p}{p \term_{\overrightarrow{j}}} \\ \\
%\\
\end{array}
\]
\caption{Structured Operational Semantics for Esterel (Part II: Transition and Termination) \label{fig::sosEsterel}}
\end{figure}

In Figure \ref{fig::sosEsterel}, the concept of termination is defined through the predicate $\term_{I, c}$, in a straightforward manner.
Exceptions are deduction rules $\DR{if4}$ and $\DR{if5}$,
which rely on (the impossibility of) the emission of the condition signal for proving termination.
In Figure \ref{fig::sosEsterel}, the deduction rules specifying a transition relation are almost identical to their counterparts in Figure \ref{fig::sosEmission}.
The most notable exceptions are deduction rules $\DR{seq0}$ to $\DR{seq4}$ and $\DR{par0}$ to $\DR{par3}$, which should consider all possible combinations of simultaneous transitions and individual transitions with (non-)termination in order to record the right target for the transition.

One advantage of our approach to the semantics of Esterel presented in \cite{Berry99,Tini00,Tini01}
is that we can capture both the logical semantics
and constructive semantics of Esterel using the same TSS
(by using two generic notions of semantics for TSS already known in the literature).
Another advantage is that it establishes a clear link between, respectively,
the logical and the constructive approaches to Esterel semantics, on the one hand
and the model- and proof-theoretic semantics of TSSs on the other hand.

\begin{defn}[Logical Semantics of Esterel\label{def::logicalSem}]
An Esterel program $p$ is logically coherent if the above given TSS is meaningful according to Semantics \ref{sem::uniqueSupported}
for $\subterms{p}$ and (predicates and) transitions labeled $\{\term_{I,p}, \emits{I, p, x}, \trans{I, p, x}\}$.
The semantics of $p$ is the set of above-mentioned predicates and transitions associated with $\subterms{p}$.
\end{defn}

Next, we show that Definition \ref{def::logicalSem} indeed satisfies the intuition behind logical coherency by re-examining the examples introduced in Section \ref{sec::cook}.

\begin{exmp}
Consider program P0, recalled below.

\begin{enumerate}
\item[P0] {\eseq{\epresent{i}{\eemit{$\mathrm{s}$}}{\enothing}}{\epresent{$\mathrm{s}$}{\enothing}{\eemit{$\mathrm{o}$}}}}
\end{enumerate}

It is straightforward to check that the following is the semantics of $P0$:

$\{P0 \emits{\{i^+\}, P0, s}$, $P0 \emits{\{i^-\}, P0, o}$, $P0 \trans{\{i^+\}, P0, i} \enothing$,
$P0 \trans{\{i^-\}, P0, o} \enothing$, $\eemit{s} \trans{\{i^+\}, P0, s} \enothing$, $\eemit{s} \trans{\{i^-\}, P0, s} \enothing$, \\
$\eemit{o} \trans{\{i^+\}, P0, s} \enothing$, $\eemit{o} \trans{\{i^-\}, P0, s} \enothing$,
$\enothing \term_{\{i^+\}, P0}$, $\enothing \term_{\{i^-\}, P0}\}$.
\footnote{For each program, we choose $\iota$, $\omega$ and $\lambda$, respectively, to comprise
only the input, output and local variables mentioned in the program at hand.
This allows us to focus only on the possibly relevant part of $I$ when considering supported models.}

Consider program P1 quoted below.

\begin{enumerate}
\item[P1] %\eencap{s}
{\epresent{s}{\eemit{s}}{\enothing}}
\end{enumerate}

It has two supported models, namely  $\{P1 \emits{\emptyset, P1, s}$, $P1 \trans{\emptyset, P1, s} \enothing$,
$\eemit{s} \emits{\emptyset, P1, s}$, $\eemit{s} \trans{\emptyset, P1, s} \enothing$, $\enothing \term_{\emptyset, P1}\}$
and $\{\eemit{s} \trans{\emptyset, P1, s} \enothing$, $\eemit{s} \emits{\emptyset, P1, s}$, $\enothing \term_{\emptyset, P1}\}$.
Hence, $P1$ is not meaningful according to Semantics \ref{sem::uniqueSupported}.

Consider program P2 recalled below.

\begin{enumerate}
\item[P2] %\eencap{s}
\epresent{s}{\enothing}{\eemit{s}}
\end{enumerate}

Program P2 does not have any supported model:  Assume, towards a contradiction,
that $P2 \emits{\emptyset, P2, s}$ is in the purported supported model of $T$.
It then follows from item 1 in Definition \ref{def::supported} that
there exists a deduction rule whose conclusion can match $P2 \emits{\emptyset, P2, s}$
and whose premises are consistent with $T$. The only candidates are $\DR{f0}$ and $\DR{f1}$;
we analyze both cases below and show that they both lead to a contradiction.

\begin{enumerate}
\item[$\DR{f0}$] The premises of the instance of $\DR{f0}$ are $P2 \emits{\emptyset, P2, s}$ and
$\enothing \emits{\emptyset, P2, s}$. It follows from item 1 of Definition \ref{def::supported}
that both predicates should be in $T$ and hence,
item 1 again applies to both predicates and in particular to $\enothing \emits{\emptyset, P2, s}$.
Hence, there should exist a deduction rule whose conclusions matches with the above predicate.
A simple syntactic check on the deduction rules of Figures \ref{fig::sosEmission} and \ref{fig::sosEsterel} reveals that none of the conclusions
can be unified with the above predicate and hence a contradiction follows.

\item[$\DR{f1}$] The premises of the instance of $\DR{f0}$ are
$\neg P2 \emits{\emptyset, P2, s}$ and $\eemit{s} \emits{\emptyset, P2, s}$, both of which should be in $T$.
Again item 1 of Definition \ref{def::supported} applies and thus, $\neg P2 \emits{\emptyset, P2, s}$  should be consistent with $T$, or in other words, $P2 \emits{\emptyset, P2, s} \notin T$, which contradicts our initial assumption.
\end{enumerate}

The next program to consider is P3, quoted below.

\begin{enumerate}
\item[P3] {\epresent{s}{\eemit{s}}{\eemit{s}}}
\end{enumerate}

Program P3 is indeed meaningful and has the following unique supported model.

$\{P3 \emits{\emptyset, P3, s}$, $P3 \trans{\emptyset, P3, s} \enothing$,
$\eemit{s} \emits{\emptyset, P3, s}$, $\eemit{s} \trans{\emptyset, P3, s} \enothing$, $\enothing \term_{\emptyset, P3}\}$.

Note that $P3 \emits{\emptyset, P3, s}$ (and/or the transition of $P3$) cannot be removed from the supported model;
to see this, it follows from item 2 of Definition \ref{def::supported} and deduction rule $\DR{e0}$ that $\eemit{s} \emits{\emptyset, P3, s} \in T$,
and following the same reasoning and deduction rule $\DR{f1}$, we have that $P3  \emits{\emptyset, P3, s} \in T$.

Program P4 is considered logically coherent but not constructive by the language designers.
Next, we show that this intuition is indeed supported by our formal definitions.

\begin{enumerate}
\item[P4] %\eencap{$\mathrm{s_0,s_1}$}
{\epar{\hspace{0.2cm} \epresent{$\mathrm{s_0}$}{\eemit{$\mathrm{s_0}$}}{\enothing}}
{\\ \hspace{0.2cm} \epresent{$\mathrm{s_0}$}{\epresent{$\mathrm{s_1}$}{\enothing}{\eemit{$\mathrm{s_1}$}}}{\enothing}} \\}
\end{enumerate}

Program P4 has a unique supported model, given below.

$\{\eemit{s_0} \emits{\emptyset, P3, s_0}$, $\eemit{s_0} \trans{\emptyset, P3, s_0} \enothing$,
$\eemit{s_1} \emits{\emptyset, P3, s_1}$, $\eemit{s_1} \trans{\emptyset, P3, s_1} \enothing$,
$\enothing \term_{\emptyset, P3}\}$.

Note that neither emission of $s_0$, nor $s_1$ cannot be present in a supported model.
First, concerning $s_1$, suppose that $s_1$ can be emitted, then it follows from item 1 of Definition \ref{def::supported}
that there should be a deduction rule supporting this emission.
This can only be due to $\DR{p1}$ and thus, the right-hand-side component of the parallel composition.
This component, in turn can only emit $s_1$ (due to deduction rules $\DR{f0}$ and then $\DR{f1}$) if $s_0$ is present and $s_1$ is absent under the same context. The latter contradicts our assumption. Similarly, suppose that the supported model contains a predicate (or transition) to the effect that $s_0$ can be emitted. We already know that no predicate for emitting $s_1$ can be in the supported model.
Hence, it follows from successive application of item 2 of Definition \ref{def::supported} using deduction rules $\DR{f0}$, $\DR{f1}$ and $\DR{e0}$ that $s_1$ can be emitted under the same context, which is already shown to lead to contradiction.

\begin{enumerate}
\item[P5] %\eencap{$\mathrm{s_0,s_1}$}
{\eseq{\epresent{$\mathrm{s_0}$}{\eemit{$\mathrm{s_1}$}}{\enothing}}{\eemit{$\mathrm{s_0}$}}}
\end{enumerate}

Program P5 is also meaningful and has a unique supported model, given below.

$\{P5 \emits{\emptyset, P5, s_0}$, $P5 \emits{\emptyset, P5, s_1}$,  $P5 \trans{\emptyset, P5, s_0} \enothing$,
$P5 \trans{\emptyset, P5, s_1} \enothing$, \\
$\epresent{\mathrm{s_0}}{\eemit{\mathrm{s_1}}}{\enothing} \emits{\emptyset, P5, s_1}$,
$\epresent{\mathrm{s_0}}{\eemit{\mathrm{s_1}}}{\enothing} \trans{\emptyset, P5, s_1} \enothing$, \\
$\eemit{s_0} \emits{\emptyset, P5, s_0}$, $\eemit{s_0} \trans{\emptyset, P3, s_0} \enothing$,
$\eemit{s_1} \emits{\emptyset, P5, s_1}$, $\eemit{s_1} \trans{\emptyset, P3, s_1} \enothing$,
$\enothing \term_{\emptyset, P5}\}$.

Note that none of the predicates or transitions concerning the emission of $s_0$ and $s_1$ can be omitted from the supported model.
If the predicate (transition) concerning the emission of $s_0$ is omitted then the first component of sequential composition terminates and hence
$s_0$ should be emitted due to the second component. Since $s_0$ should always be emitted, the emission of $s_1$ is guaranteed by the first component of sequential composition.

\end{exmp}

\begin{defn}[Constructive Semantics of Esterel]
An Esterel program $p$ is constructive if for each signal $s$ and each input evaluation $I$ either $p \emits{I, p, s}$  and  $p \trans{I, p, s} p'$ (for some $p'$) or
$\neg p \emits{I, p, s}$ and $p \ntrans{I, c, s}$ has a supported proof and moreover, either $p \term_{I,p}$ or $\neg p \term_{I,p}$ has a supported proof.
\end{defn}

To illustrate this semantics and identify its differences with the logical semantics,
we reconsider those programs whom are considered non-constructive but logically coherent in Section \ref{sec::cook}.

\begin{exmp}
Consider program P3. This program is both intuitively and formally shown to be logically coherent.
Moreover, in Section \ref{sec::cook}, we introduced this program as a canonical example of a non-constructive program.
Next, we show that it is also formally non-constructive since neither
$P3 \emits{\emptyset, P3, s}$ nor $\neg P3 \emits{\emptyset, P3, s}$ have a supported proof (a similar reasoning shows
that neither $P3 \trans{\emptyset, P3, s} p'$  for any $p'$ nor $P3 \ntrans{\emptyset, P3, s}$ have a supported proof).
Suppose $P3 \emits{\emptyset, P3, s}$ has a supported proof, then its proof is either due to $\DR{f0}$ or $\DR{f1}$.
In the former case, the nodes placed above our proof obligation are $P3 \emits{I, P3, s}$ and $\eemit{s} \emits{\emptyset, P3, s}$. While the latter
has a supported proof (due to $\DR{e0}$), the former was our original proof obligation, thus, it only remains to check the alternative option due to $\DR{f1}$.
The premises of $\DR{f1}$ are then $\neg P3 \emits{I, P3, s}$ and $\eemit{s} \emits{\emptyset, P3, s}$. Again the latter formula has a supported proof but
the former is the negation of our proof obligation and thanks to Theorem \ref{th::supportedConsistency}, we know that if  $\neg P3 \emits{I, P3, s}$  has a supported proof then $P3 \emits{I, P3, s}$ cannot have a supported proof. Similarly, if $\neg P3 \emits{I, P3, s}$  has a supported proof, then a negation of a premise of all deduction rules that can match $P3 \emits{I, P3, s}$  must have a supported proof. These two rules are again $\DR{f0}$ and $\DR{f1}$.
The negation of the common premise of these two rules, i.e.,  $\eemit{s} \emits{\emptyset, P3, s}$ cannot have a supported proof (following Theorem  \ref{th::supportedConsistency}, because the premise itself has a supported proof). Hence a negation of both
$P3 \emits{I, P3, s}$  and $\neg P3 \emits{I, P3, s}$ should have supported proofs, which is again impossible due to Theorem \ref{th::supportedConsistency}.

Program P4 is not constructive since neither $P4 \emits{\emptyset, P4, s_0}$, nor its negation have a supported proof.
The only possible proof for the emission predicate can be due to $\DR{p0}$ or $\DR{p1}$.
The case for $\DR{p1}$ does not lead to a supported proof since the right-hand-side does not contain any emit statement for $s_0$.
If the supported proof is due to $\DR{p0}$, then it should hold that $P4 \emits{\emptyset, P4, s_0}$ which was to be proven.
The negation of the predicate, i.e., $P4 \emits{\emptyset, P4, s_0}$ does not have a supported proof, either.
Since then a negation of a premise of $\DR{p0}$ and $\DR{p1}$ should have a supported proof.
The negation of the only premise of $\DR{p0}$ is $\epresent{\mathrm{s_0}}{\eemit{\mathrm{s_0}}}{\enothing}$ $\emits{\emptyset, P4, s_0}$, which in turn means that
a negation of a premise of $\DR{f0}$ or $\DR{f1}$ must have a supported proof.
Consider $\DR{f0}$, its two premises are $P4 \emits{\emptyset, P4, s_0}$, but we were seeking a proof of its negation and
$\eemit{s_0} \emits{\emptyset, P4, s_0}$, whose negation cannot be proven.

Program P5 is not constructive, either. We next show that neither $P5 \emits{\emptyset, P5, s_0}$  nor its negation are provable.
The purported supported proof for predicate $P5 \emits{\emptyset, P5, s_0}$ is due to one of the rules $\DR{s0}$ to $\DR{s2}$.
Next, we analyze each case and show that it leads to a contradiction.

\begin{enumerate}
\item[$\DR{s0}$] Then, it should hold that $\epresent{\mathrm{s_0}}{\eemit{\mathrm{s_1}}}{\enothing} \emits{\emptyset, P5, s_0}$. This, in turn, can be either
due to $\DR{f0}$ or $\DR{f1}$. If the predicate is due to $\DR{f0}$, then we should have a supported proof for $P5 \emits{\emptyset, P5, s_0}$, which was to be proven.
If the proof is due to $\DR{f1}$, then $\neg P5 \emits{\emptyset, P5, s_0}$ should have a supported proof, which is impossible due to Theorem \ref{th::supportedConsistency}.

\item[$\DR{s1}$] Then, it should hold that $\epresent{\mathrm{s_0}}{\eemit{\mathrm{s_1}}}{\enothing} \term_{\emptyset, P5}$. This termination can be due to either
$\DR{if4}$ or $\DR{if5}$. None of these two are possible since otherwise, respectively, $P5 \emits{\emptyset, P5, s_0}$ or $\neg P5 \emits{\emptyset, P5, s_0}$ should have a supported proof.

\item[$\DR{s2}$] Then, it should hold that $\epresent{\mathrm{s_0}}{\eemit{\mathrm{s_1}}}{\enothing} \term_{\emptyset, P5, s'} p'$ for some $s'$ and $p'$. This transition is due to either $\DR{if0}$ or $\DR{if1}$. Again, both cases lead to a contradiction due to a similar reasoning as in item $\DR{s0}$.
\end{enumerate}

\end{exmp}

As a side note, the common intuition and the similarities between deduction rules of Figures \ref{fig::sosEmission} and \ref{fig::sosEsterel} may suggest that we can replace deduction rules of Figure \ref{fig::sosEmission} with the following rule (or even do without the emission predicates and make the same changes in the deduction rule for if-then-else statements in Figure \ref{fig::sosEsterel}):

\[
\DR{emit} \sosrule{p \trans{I, c,x} p'}{p \emits{I, c,x}}
\]

This change leads to a much more restrictive semantics, which is unable to provide supported proofs for transitions of perfectly acceptable programs such as the following:

\begin{enumerate}
\item[P6] %\eencap{$\mathrm{s_0,s_1}$}
{\epar{\epresent{$\mathrm{s}$}{\eemit{$\mathrm{o}$}}{\enothing}}{\eemit{$\mathrm{s}$}}}
\end{enumerate}

To see this, the reader may try to prove that P6 can emit signal $o$ using deduction rule $\DR{par0}$. The proof of the premise of $\DR{par0}$ then should rely on $\DR{if0}$ and hence due to deduction rule $\DR{emit}$, we need to prove that  $s$ can be emitted (for the if-then-else to be able to take a transition). In turn,  this can only be due to $\DR{par1}$. But to apply $\DR{par1}$, we need to know that the left-hand-side component can take a transition (in order to record its target), which is what we wanted to prove initially. This cycle is broken in our semantics, by deduction rule $\DR{p1}$ which only considers one of the two components to infer the emission of $s_0$ (without trying to record the target of the transition). The following proof illustrates why this program is indeed constructive.

\[
\sosrule{
\sosrule{
\sosrule{
\sosrule{}{
\eemit{\mathrm{s}} \emits{\emptyset, P6, s}}
}{
P6 \emits{\emptyset, P6, s}
}
\quad
\sosrule{}{\eemit{\mathrm{o}} \trans{\emptyset, P6, o} \enothing}
}
{
\epresent{\mathrm{s}}{\eemit{\mathrm{o}}}{\enothing} \trans{\emptyset, P6, o} \enothing
}
\quad
\sosrule{}{\eemit{\mathrm{s}} \trans{\emptyset, P6, s} \enothing}
}{P6 \trans{\emptyset, P6, o} \epar{\enothing}{\enothing}}
\]

\section{Conclusions and Future Work\label{sec::conc}}
In this paper, we presented a link between the intuitive notions of logical coherency and
constructiveness in the semantics of Esterel on the one hand, and the formal notions of supported models and supported proofs
in the semantics of Structured Operational Semantics, on the other hand.
By means of several canonical examples from the literature, we showed that our formal definitions indeed capture the intuitive
criteria put forward by the language designers.

Several formalizations of these two intuitive criteria exist in the literature. For example \cite{Berry99,Berry07} present three formalizations of
constructive semantics of Esterel. In \cite{Tini00,Tini01} another formalization of constructive semantics of Esterel is presented and is proven to
coincide with one of the notions in \cite{Berry99}. A rigorous comparison between all these notions and the ones presented in this paper remains as a topic for future research.

In the semantics presented in this paper, we abstracted from the issues of exceptions (traps), loops  and time.
We expect that one can include these aspects without any substantial change in the semantics presented in this paper using the modular semantics approach of \cite{Mosses04a,Mosses08}. This remains as another interesting exercise for the future.

\paragraph{Acknowledgements.} Inspiring discussions with Jean-Pierre Talpin and Paul Guernic are gratefully acknowledged. The author would like to thank the anonymous reviewers of SOS 2009 for their insightful reviews.

\end{document}